# The Impact of Vocalization Loudness on COVID-19 Transmission in Indoor Spaces

Santiago Barreda[1], Sima Asadi[2†], Christopher D. Cappa[3], Anthony S. Wexler[3,4,5,6],
Nicole M. Bouvier[7,8], and William D. Ristenpart[2*]

[1] Dept. of Linguistics, Univ. of California Davis, 1 Shields Ave., Davis, CA 95616 USA.
[2] Dept. of Chemical Engineering, University of California Davis, 1 Shields Ave., Davis, CA 95616 USA.
[3] Dept. of Civil & Environmental Engineering, Univ. of California Davis, 1 Shields Ave., Davis, CA 95616 USA
[4] Dept. of Mechanical & Aerospace Engineering, Univ. of California Davis, 1 Shields Ave., Davis, CA 95616 USA.
[5] Air Quality Research Center, Univ. of California Davis, 1 Shields Ave., Davis, CA 95616 USA.
[6] Dept. of Land, Air and Water Resources, Univ. of California Davis, 1 Shields Ave., Davis, California 95616 USA.
[7] Dept. of Medicine, Div. of Infectious Diseases, Icahn School of Medicine at Mount Sinai, 1 Gustave Levy Place, New York, NY 10029 USA.
[8] Dept. Microbiology, Icahn School of Medicine at Mount Sinai, 1 Gustave Levy Place, New York, NY 10029 USA.

† Present address: Dept. of Civil & Environmental Engineering, Massachusetts Institute of Technology, Cambridge, MA 01239 USA.

*Corresponding author: William D. Ristenpart, Dept. of Chemical Engineering, Univ. of California Davis, 1 Shields Ave., Davis, CA 95616 USA, (530) 752-8780, wdristenpart@ucdavis.edu

## Abstract

There have been several documented outbreaks of COVID-19 associated with vocalization, either by speech or by singing, in indoor confined spaces. Here, we model the risk of in-room airborne disease transmission via expiratory particle emission versus the average loudness of vocalization and for variable room ventilation rates. The model indicates that a 6-decibel reduction in average vocalization intensity yields a reduction in aerosol transmission probability equivalent to doubling the room ventilation rate. The results suggest that public health authorities should consider implementing "quiet zones" in high-risk indoor environments, such as hospital waiting rooms or dining facilities, to mitigate transmission of COVID-19 and other airborne respiratory diseases.

**Main Text**

There is an emerging consensus that COVID-19 is transmissible via airborne aerosol particles that are emitted when infected individuals breathe, speak, sneeze, or cough [1-8]. The relative contributions of these expiratory activities to airborne transmission remains unclear, but multiple outbreaks have been documented in which asymptomatic carriers were speaking or singing in confined indoor spaces with susceptible individuals [9,10]. Vocalization causes micron-scale droplets of respiratory mucosa to form via a "fluid-film-burst" mechanism, either in the lungs during inhalation due to expansion of the alveoli, or in the vocal cords due to rapid opening and closing of the glottis during phonation [11-13]. Upon exhalation into the ambient air these droplets rapidly evaporate to leave behind dried aerosol particles large enough to carry viable virus that, although too small to see by eye, are lightweight enough to remain suspended for long times; particles smaller than approximately 5 µm will typically be removed from rooms by air exchange rather than gravitational settling [14-16]. Expiratory particles in this size range from exhaled breath are known to carry infectious influenza virus [17]; likewise, viable SARS-CoV-2, the virus responsible for COVID-19, has been observed in micron-scale aerosol particles sampled from hospital air several meters away from infected patients [18].

We recently demonstrated that the emission rate of micron-scale respiratory aerosol particles strongly correlates with the loudness of speech [19,20]. An increase in vocalization intensity of about 35 decibels, roughly the difference between whispering and shouting, yields a factor of 50 increase in the particle emission rate. We also reported that the size distribution of the dried particles is independent of vocalization loudness, and that certain individuals, for unclear reasons, act as superemitters during vocalization, releasing an order of magnitude more particles than average. We hypothesized that airborne disease transmission might occur more readily in noisy environments where infected individuals must speak loudly, thus causing enhanced emission of infectious expiratory particles into the air [19]. Epidemiologists have speculated that recent COVID-19 outbreaks in churches [9], bars [21], or meat processing facilities [22,23] might be due in part to the loudness of these environments. In response, various public health authorities have provided official recommendations that discourage [24-27] or even explicitly prohibit [28] singing and other loud vocalizations, or prohibit conditions like playing loud music that necessitate raising of voices [29].

Much remains unknown, however, about the possible link between vocalization loudness and airborne disease transmission. If virus-laden particles are emitted via vocalization, and if louder vocalization yields more particles, then a key question is: how does the loudness of vocalization affect the transmission probability?

As a starting point to addressing this question, we use the simplest quantitative theoretical model for airborne disease transmission, named the Wells-Riley model after the early investigators who performed this pioneering work [30,31]. Detailed derivations and assessments of the accuracy of the Wells-Riley model are provided elsewhere [32,33]; here we simply use the model framework, which is that the transmission probability follows the complement of a Poisson distribution,

$$P = 1 - e^{-\mu}, \tag{1}$$

where $\mu$ is the expected number of infectious pathogens that a susceptible individual inhales. This probability distribution assumes that only one pathogen is necessary to initiate infection, but more complicated expressions are available to account for larger minimum infectious doses [14]. In the classic Wells-Riley formulation, $\mu$ is calculated with the assumption that pathogens are emitted at a rate $q$ pathogens per second from one or more infected individuals in a room with instantaneously well-mixed air, so that the relative positions of the infected and susceptible individuals are irrelevant. As such, the model does not account for potential enhanced transmission by direct inhalation of the respiratory plume emitted by an infected individual, but the assumption of well-mixed air serves in part as the basis for minimum ventilation standards promulgated by CDC [34] and ASHRAE [35] because it yields a lower bound for transmission risk to all room occupants regardless of position. The Wells-Riley model further assumes that the room has a ventilation rate of $Q$ liters per minute delivering fresh (pathogen-free) air, and that susceptible individuals are moving $B$ liters of air in and out of their lungs per minute of breathing (i.e., the minute ventilation). In the case where there is just one infected individual, the expected value is

$$\mu = \frac{\eta q B}{Q} t, \tag{2}$$

where $t$ is the total exposure time. The parameter $\eta$ here represents an infection efficiency ($0 < \eta < 1$) that includes physical effects, like the deposition efficiency within the respiratory tract of the susceptible individual, and immunological effects, like the ability of the immune system to repress the infection. For a minimum infectious dose of 1 pathogen, the quantity $\eta q$ is equivalent to the "quanta" of infectivity initially used by Wells and Riley in their models [30,31].

It is already well known from equations (1) and (2) that increasing the exposure time or decreasing the room ventilation rate will increase the expected number of inhaled pathogens and the corresponding transmission probability [36]. What is new here is consideration of the impact of vocalization intensity on the virus aerosolization rate $q$. The particle emission rates that we previously reported were measured in a laboratory environment while using a microphone and decibel meter placed near the mouth [19,20]. Importantly, the particle emission rate varied linearly with the root-mean-square amplitude as measured by the microphone; the amplitude varies nonlinearly with the corresponding sound pressure level in decibels (Fig. S1). Using these measurements, we can relate expected particle emission rates to different sound pressure levels, measured in C-weighted decibels (dBC). Full details are presented in the Supplementary material; the final result is that the average particle emission rate is estimated to depend on the vocalization intensity $L_{p1}$, measured in dBC at 1 m from a non-masked speaker, as

$$N_{avg} = (1 - \phi) \, \widehat{N}_{br} + \phi \, \widehat{N}_{voc} \left( \frac{L_{p1} + 25}{105} \right)^{10.6}, \tag{3}$$

where $\widehat{N}_{br}$ and $\widehat{N}_{voc}$ are scaled expiratory particle emission rates for breathing and vocalization, respectively, that depend on the expiratory flowrates. The parameter $\phi$ represents the fraction of time the infected individual is vocalizing during the exposure time; $\phi$ is close to zero for individuals who vocalize rarely such that breathing-related emission dominates, and approaches one for those who vocalize continuously, such as in singing or chanting. The average virus aerosolization rate then is

$$q = C_v V_d \left[ \xi (1 - \phi) \, \widehat{N}_{br} + \phi \, \widehat{N}_{voc} \left( \frac{L_{p1} + 25}{105} \right)^{10.6} \right], \tag{4}$$

where $C_v$ is the viral concentration in the respiratory fluid of the infected individual, and $V_d$ is the pre-evaporation volume of droplets emitted during vocalization. The parameter $\xi = V_b / V_d \approx 0.5$ is the volume ratio of droplets emitted via breathing versus vocalization; several researchers have found that vocalization yields significantly larger droplets than breathing [11,13,19]. Combination of equations (2) – (4) into (1), and noting that the ventilation rate in a room with volume $V_{room}$ is related to the air changes per hour as $Q = V_{room} ACH$, yields the desired probability,

$$P = 1 - \exp\left(-k \frac{\left[\xi(1-\phi)\,\bar{N}_{br} + \phi\,\bar{N}_{voc}\left(\frac{L_{p1}+25}{105}\right)^{10.6}\right]}{ACH} t\right). \tag{5}$$

Here $k = \frac{\eta B C_v V_d}{V_{room}}$ is an effective rate constant composed of parameters that, for a given room and specific virus, are not readily alterable by human interventions.

The striking feature of equation (5) is the large power-law dependence on the vocalization intensity. A contour plot of the transmission probability versus vocalization intensity and duration illustrates this pronounced impact for a 1-hour exposure time in a room with three ACH (Figure 1). The transmission probability is lowest in the bottom left corner, corresponding to infectors who vocalize rarely and quietly, as might be observed in a library or quiet office space. In contrast, the transmission probability increases gradually with duration and rapidly with intensity. It reaches maximal values in the top right corner, corresponding to infectors who vocalize loudly and close to continuously, as might be observed in a noisy bar environment or at a choir practice.

The model also gives insight on the cost-benefit analysis of increasing the room ventilation rate. Fig. 2A shows the transmission probability versus vocalization intensity for different ACH values. As expected, doubling the ventilation rate of fresh (pathogen-free) air decreases the transmission probability. A notable feature, however, is that a similar reduction in transmission probability can be gained, without changing the ventilation rate, simply by decreasing the vocalization intensity by approximately 6 dBC. This reduction can be quantified via a risk reduction factor,

$$f = \frac{P_{original} - P_{intervention}}{P_{original}}, \tag{6}$$

where $P_{original}$ is the probability at some initial condition and $P_{intervention}$ is the adjusted probability via an intervention either with an increased ventilation rate or decreased vocalization.

For simplicity, we can focus on small values of $\mu$ such that asymptotically $P \approx \mu$, in which case the risk reduction factor for doubling the room ventilation rate is $f = \frac{1}{2}$. If the infected individual simply vocalizes half as often (i.e., $\varphi$ is halved), then to good approximation $f \approx \frac{1}{2}$ as well. Furthermore, keeping the room ventilation rate and the vocalization duration fixed, the risk reduction factor for decreasing the vocalization intensity by $\delta$ decibels is

$$f = 1 - \left(\frac{L_{p1} + 25 - \delta}{L_{p1} + 25}\right)^{10.6}. \tag{7}$$

To achieve a 50% reduction in risk for vocalization that ordinarily would occur at 60 dBC would require a decrease of only $\delta = 5.4$ dBC. More precise calculations of the risk reduction factor (Fig. 2B) show that in general, a 10 dBC decrease in average vocalization intensity is always more effective at reducing risk of aerosol transmission than doubling the ventilation rate.

The risk reduction achieved either by increasing room ventilation or by decreasing the loudness of vocalization is insensitive to the pathogen concentration in respiratory emissions or their infection efficiency, though those quantities do affect the actual probability of transmission. In other words, the numerical values of the probabilities shown in Figs 1 and 2A will vary with the viral load of the infector, but the overall shape of the curves will remain the same. Similarly, wearing of masks will reduce the particle emission rate of the infector and decrease the effective deposition efficiency in susceptible individuals and thus decrease the overall probability, but the relative risk reduction as characterized here will remain unchanged. We also emphasize that the Wells-Riley model explicitly assumes the air is well mixed, and that more sophisticated plume or puff models [37,38] or computational fluid dynamics models [39,40] are required to account for the directionality and turbulent diffusivity of the airflow and proximity of individuals. Whatever transport model is used, however, the vocalization source terms in equations (3) and (4) suggest that reductions in vocalization intensity will strongly decrease the amount of virus available to be transported, and thus decrease the overall transmission probability.

To relate these proposed decibel decreases to real-world situations, we consider typical noise levels in different indoor environments, often measured in A-weighted decibels (dBA), which are thought to better reflect subjective perceptions of loudness. Ambient noise in restaurants is typically between 65–80 dBA, with an average of 73 dBA [41], and background noise levels of 75 dBA have been observed at day-care facilities [42]. Music plus crowd noise in bars and nightclubs can average as high as 90-100 dBA [43]. The relationship between ambient noise levels and the speech loudness necessary for comprehension is complex, but in general speech must be nearly the level of the background noise to be understood, and speakers adjust their vocalization intensity to maintain a positive signal-to noise ratio when possible [44-46]. As a result, all other things being equal, a reduction in background noise on the order of 5-10 decibels will facilitate, if not directly result in, a corresponding reduction in average speaking levels. Further, the relatively high amount of background noise in many public spaces suggests that there is considerable room to reduce noise levels behaviorally (e.g., turning music down, encouraging silence), since noise is not inherent to the operation of many of these spaces (as opposed to industrial facilities). When wearing facemasks, the reduction in the background noise necessary to achieve a similar magnitude reduction in transmission risk may be larger owing to the need to speak more loudly through the mask [47]. A more detailed analysis of mask filtration efficacy and vocalization through masks is necessary to characterize the impact of this effect on transmission probability.

There are tremendous installation, maintenance, and energy costs associated with increased ventilation rates, especially in air conditioned or heated indoor spaces [48]. In practice many ventilation systems recycle a substantial fraction of the room air, so increasing the flow rate of fresh (pathogen-free) air requires even more ACH.  In comparison, there is little cost for signage and dissemination campaigns aimed at discouraging use of loud voices in shared indoor environments.  Libraries, for example, are traditionally quiet in part because librarians promulgate social conventions against loud conversations. The results presented here suggest that public health authorities should consider fostering comparable social conventions in hospital waiting rooms or other high-risk environments where people must congregate and social distancing is difficult to maintain. The results also suggest that epidemiologists should consider the acoustic conditions of indoor environments as a potential contributing factor in situations where outbreaks of COVID-19 or other viral respiratory diseases might occur.

**Acknowledgments**

We thank the National Institute of Allergy and Infectious Diseases of the National Institutes of Health (NIAID/NIH), grant R01 AI110703, for supporting this research.

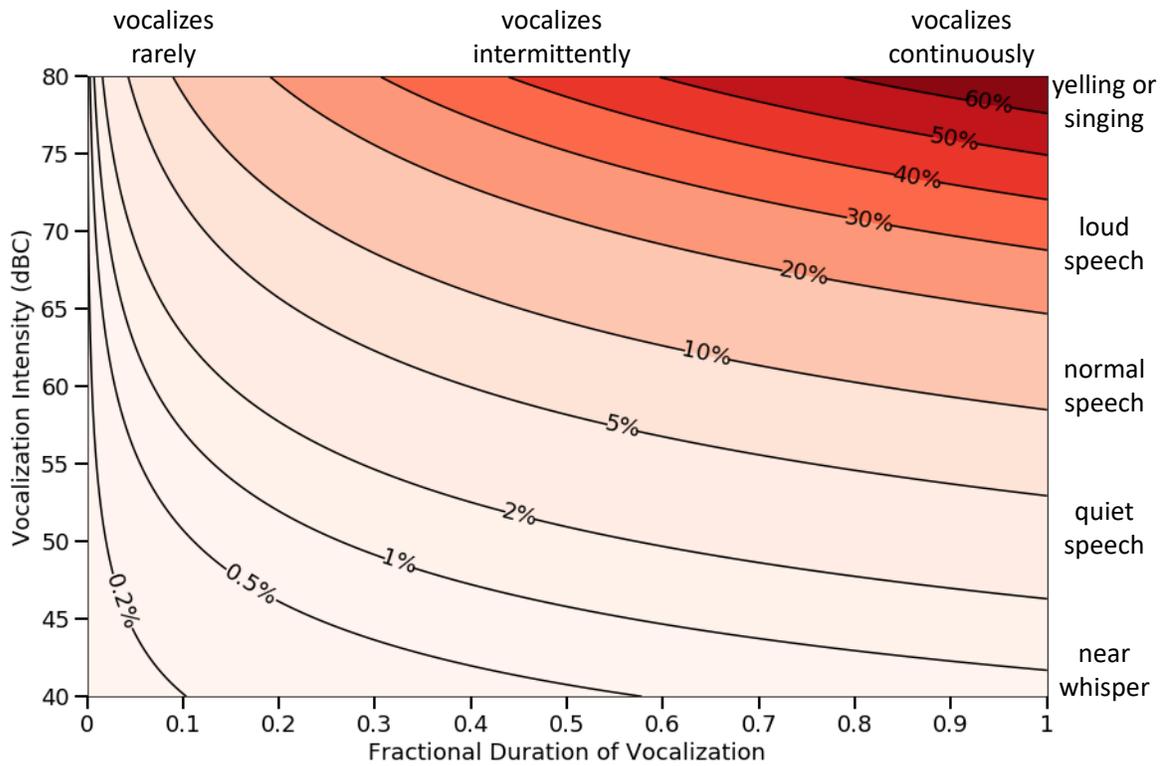

**Figure 1** – Contour plot of transmission probability for 1 hour of exposure to a vocalizing individual infected with SARS-CoV-2, in a room with 3 ACH, versus the vocalization loudness (measured at 1 meter) and the fractional duration of vocalization ($\phi$) by the infector during the hour-long exposure. Model parameters are listed in Table S1.

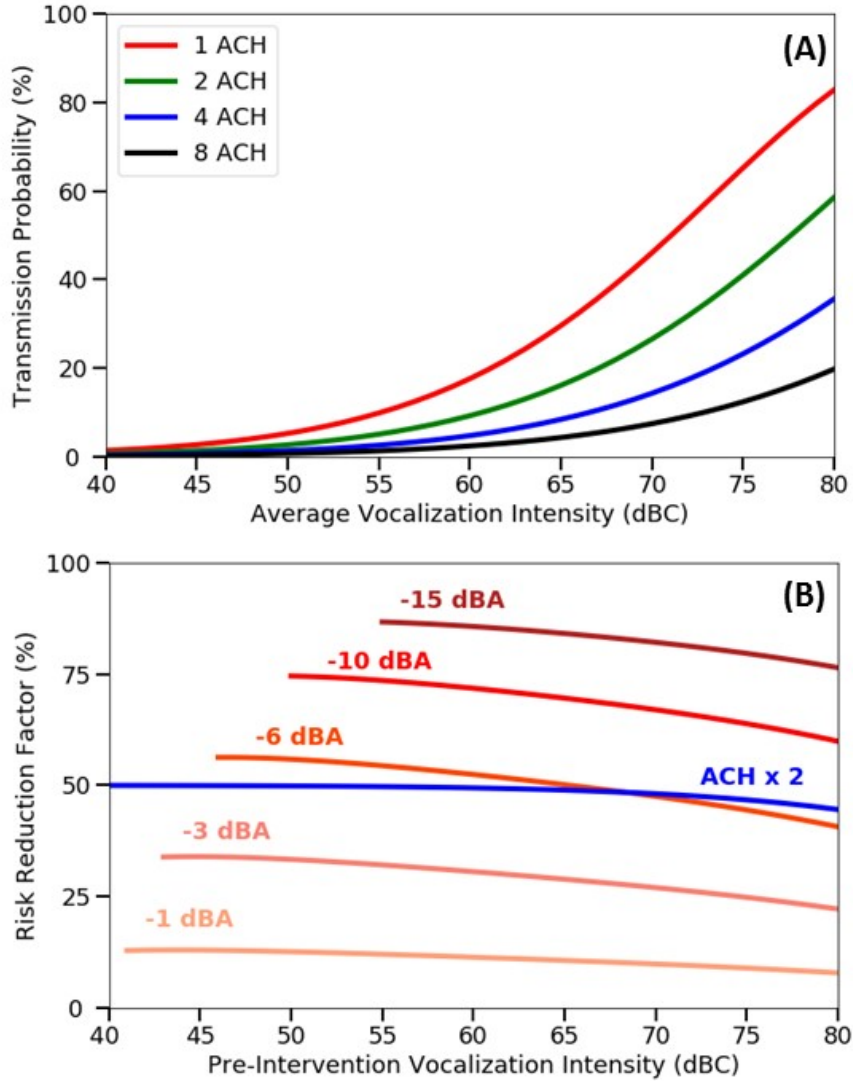

**Figure 2** – **(A)** Probability of susceptible individuals becoming infected with SARS-CoV-2 after 1-hour of exposure, during which infector vocalized half of the time ($\phi = 0.5$) at the specified sound pressure level (measured 1 meter from the speaker). **(B)** The risk reduction factor versus original vocalization intensity for different decreases in vocalization intensity (red curves) or increasing the ventilation by a factor of two (blue curve). Model parameters listed in Table S1.

# Supplementary Information:

# The Impact of Vocalization Loudness on COVID-19 Transmission in Indoor Spaces

Santiago Barreda, Sima Asadi, Chris Cappa, Anthony S. Wexler,

Nicole M. Bouvier, and William D. Ristenpart

Here we derive equation (3) in the main text, which describes the relationship between the measured vocalization intensity, as measured in decibels, and the average emission rate of expiratory aerosol particles. The empirical data and experimental methods are described in detail by Asadi et al., *Scientific Reports* 2019; for reference similar results were reported by Asadi et al., *PLoS One* 2020. In brief, participants either breathed or vocalized into a funnel connected to an aerodynamic particle sizer (APS) placed in a HEPA-filtered laminar flow (Fig. S1a). The APS draws in 5 liters/min of air from the funnel, of which 80% comprised a sheath flow and 20% a sample flow ($F_s$) measured in the detector. A microphone and a decibel meter placed near the funnel entrance simultaneously measured the root-mean-square amplitude, $A_{rms}$, of the vocalization and the corresponding sound pressure level (SPL) in C-weighted decibels.

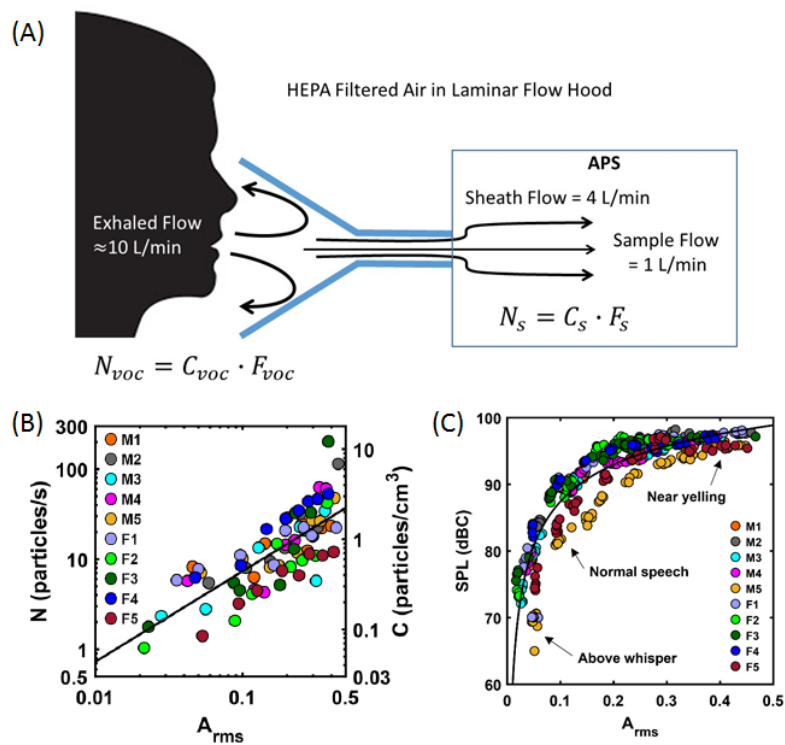

**Figure S1** – **(A)** Schematic of the experimental apparatus (not to scale) and approximate airflow streamlines. Microphone and decibel meter next to the funnel are not shown. See also Asadi *et al.* 2019 supplementary Fig 1 and Fig S12. **(B)** Scatter plot of the particle emission rate detected in the APS versus the vocalization amplitude. Solid line has a slope of 1.004. Reproduced from Fig. 2c of Asadi *et al.* 2019. **(C)** Calibration curve relating the amplitude to the sound pressure level, measured at 5 cm from the mouth, in C-weighted decibels. Solid line is the power-law fit given by equation S5. Reproduced from Fig. S1 of Asadi *et al.* 2019.

The key finding, shown in Fig. S1b, is that the rate of particles moving through the detector in the sample flow, in particles per second (p/s), varied linearly with the vocalization amplitude,

$$N_s = \kappa\, A_{rms} .$$ (S1)

The amplitude varied from 0 to 0.5 (arbitrary units), and the slope $\kappa$ was approximately 30 to 40 particles per second for speaking or 'singing' respectively (cf. Figs 2c and 3b of Asadi et al. 2019). Importantly, however, not all of the exhaled air was fed into the detector. Typical exhalation flow rates during breathing and vocalization ($F_{voc}$) range from 8 to 12 L/min (Loudon 1988, Gupta et al. 2010), while the APS only detected particles in the sample flow at 1 L/min. As the breathing and vocalization flow rates exceed the total APS flow rate (5 L/min) there is no dilution of the sampled air. Thus, to estimate the total particle emission rate, we equate the concentration in the detector to the exhaled concentration in the funnel ($C_s = C_{voc}$), yielding the relationship

$$N_{voc} = \frac{F_{voc}}{F_s} N_{s,voc},$$ (S2)

where $N_{voc}$ is the total particle emission rate from vocalization (p/s). A similar statement pertains to the (non-vocalization) particle emission rate during breathing, $N_{br}$. Over sufficiently long time periods, the average total particle emission rate will reflect the relative amounts of time spent breathing versus vocalizing, viz.,

$$N_{avg} = (1 - \phi)N_{br} + \phi\, N_{voc},$$ (S3)

where $0 \leq \phi < 1$ is the fraction of time the individual spends vocalizing. Inserting the relationships defined in (1) and (2) into (3) yields

$$N_{avg} = (1 - \phi)\frac{F_{br}}{F_s}N_{s,br} + \phi\kappa\frac{F_{voc}}{F_s}A_{rms}.$$ (S4)

Next, we note that the microphone amplitude $A_{rms}$ is related to the sound pressure level in decibels via a power-law relationship of the form

$$L_{p0} = cA_{rms}^b,$$ (S5)

as shown in Fig. S1C. Nonlinear regression yields best fit values of $b = 0.094$ and $c = 105$ dBC. The decibel readings were recorded 6.5 cm from the mouth, but it is standard to report sound pressure levels at a distance of 1 m from the noise source. Accordingly, we adjust the sound pressure level as

$$L_{p1} = L_{p0} + 20\log_{10}\frac{r_0}{r_1} = L_{p0} - \Delta,$$ (S6)

where $\Delta = 25$ dBC for $r_1 = 1$ m. Combination of (1), (2), (5) and (6) yields the particle emission rate versus sound pressure level,

$$N_{voc} = \kappa \frac{F_{voc}}{F_s} \left( \frac{L_{p1} + \Delta}{a} \right)^{1/b}. \tag{S7}$$

Finally, combining everything into equation S4 yields the desired expression,

$$N_{avg} = (1 - \phi) \frac{F_{br}}{F_s} N_{s,br} + \phi \kappa \frac{F_{voc}}{F_s} \left( \frac{L_{p1} + \Delta}{a} \right)^{1/b}. \tag{S7}$$

For convenience we define $\hat{N}_{br} = N_{s,br} \frac{F_{br}}{F_s}$ and $\hat{N}_{voc} = \kappa \frac{F_{voc}}{F_s}$, and substitution of the empirical coefficients $a$, $b$, and $\Delta$ yields equation (3) in the main text.

The independent variables of interest in equation S7 for modeling the transmission probability are $\phi$ and $L_{p1}$. All other parameters are known from the empirical measurements reported by Asadi et al., except for the expiratory flowrates $F_{br}$ and $F_{voc}$. As noted by several authors, the relationship between measured sound pressure level and the expiratory flow rate is quite complicated, and depends on the pitch (fundamental frequency), the "open quotient" of the vocal cords, the lung pressure and vocalization pressure threshold, and the glottal and epiglottal resistances (Schneider and Baken 1984, Titze 1992, Jiang et al. 2016). As first summarized succinctly by Rubin et al., there is a "lack of any consistent relationship between sound pressure levels and air flow" (Rubin et al. 1967). Accordingly, as a first approximation here we simply treat the average flow rate during vocalization as a fixed constant independent of the sound pressure level, which in general will yield a conservative underestimate of the total particle emission rate as sound pressure level increases. Model parameters and sources are listed in Table S1.

| Parameter | Value | Reference |
|:---:|:---:|:---:|
| $V_{room}$ | 300 m$^3$ | – |
| $t$ | 1 hour | – |
| $B$ | $1.3 \times 10^{-4}$ m$^3$/s | Chen et al. |
| $C_v$ | $10^8$ virions/mL | To et al. |
| $\eta$ | 0.4 | Rissler et al. |
| $F_{br}$ | 8 L/min | Gupta et al. |
| $F_{voc}$ | 10 L/min | Gupta et al. |
| $F_s$ | 1 L/min | Asadi et al. |
| $N_{br}$ | 0.05 particles/s | Asadi et al. |
| $\kappa$ | 40 particles/s | Asadi et al. |
| $\theta$ | 0.32 | Liu et al. |
| $\xi$ | 0.51 | Asadi et al. |
| $V_d$ | 0.18 pL | Asadi et al. |

**Table S1** – Parameter models used in Figs 1 and 2 in the main text.

**References in Supplementary**